\documentclass[10pt]{iopart}
\usepackage{graphicx} 
\usepackage{iopams}
\usepackage[ansinew]{inputenc}
\usepackage[usenames]{color}
\newcommand{\bra}[1]{\langle{#1}|}
\newcommand{\ket}[1]{|{#1}\rangle}

\begin{document}
\title{Prime Factorization of Arbitrary Integers with a Logarithmic Energy Spectrum}
\author{F Gleisberg$^{1}$, F Di Pumpo$^1$, G Wolff$^1$ and 
W P Schleich$^{1,2}$ 
} 
%
%
\address{$^{1}$~{Institut f\"ur Quantenphysik and Center for Integrated Quantum Science and Technology ($\rm {IQ}^{\rm {ST}}$), Universit\"at Ulm, D--89069 Ulm, Germany}} 
\address{$^{2}$~Hagler Institute for Advanced Study at Texas A \& M University, Texas A \& M AgriLife Research, 
Institute for Quantum Studies and Engineering (IQSE) and
Department of Physics and Astronomy, Texas A \& M University,
College Station, Texas 77843-4242, USA}

\ead{ferdinand.gleisberg@alumni.uni-ulm.de}

\begin{abstract}
We propose an iterative scheme to factor numbers based on the quantum
dynamics of an ensemble of interacting bosonic atoms stored in a trap where the single-particle
energy spectrum depends logarithmically on the quantum number.
When excited by a time-dependent
interaction these atoms perform Rabi oscillations between the ground state
and an energy state characteristic of the factors.
The number to be factored is encoded into the frequency
of the sinusoidally modulated interaction.
We show that a measurement of the energy of 
the atoms at a time chosen at random yields the factors with probability one half. 
We conclude by discussing a protocol to obtain
the desired prime factors employing a logarithmic energy spectrum
which consists of prime numbers only.
\end{abstract}


\begin{center}
	\vspace{2em}{
	\hspace{6.5em}{
	\begin{minipage}[c]{0.8\textwidth}
		This Accepted Manuscript is available for reuse under a CC BY-NC-ND 3.0 licence after the 12 month embargo period provided that all the terms of the licence are adhered to.
		
		This is a peer-reviewed, un-copyedited version of an article published in Journal of Physics B: Atomic, Molecular and Optical Physics. IOP Publishing Ltd is not responsible for any errors or omissions in this version of the manuscript or any version derived from it. The Version of Record is available online at 10.1088/1361-6455/aa9957.
	\end{minipage}}}
\end{center}


\maketitle

\section{Introduction}\label{Intro}

The decomposition of a positive integer into a product
of prime factors is a difficult problem 
in number theory. The hardest one is to factor
semiprimes, {\it i.e.} numbers composed of two primes, provided
they are different and quite large. No wonder that this
property of semiprimes finds applications in cryptography.

For example, the famous RSA algorithm \cite{RSA} uses a public key in
order to code a message. However, only the owner
of the private key can quickly decode it because among the steps needed to achieve this
secret key from the public one is the factorization of a very large
semiprime. All known factoring algorithms run in non-polynomial time on classical 
computers thus prohibiting unauthorized decoding in a reasonable time.
But on a large ideal quantum computer Shor's celebrated 
factorizing algorithm \cite{Shor} takes only polynomial time and is therefore expected 
to break the RSA scheme in the future.

We have studied \cite{Gleisberg,Gleisberg20152556} the factorization of semiprimes
using one- and two-dimensional traps\footnote{Such traps may be realized in the future using methods
from the emerging field of atomtronics \cite{Boshier}
as {\it e.g.} the technique
of ''painted potentials'' \cite{Henderson} or ''adiabatic potentials'' \cite{garraway2016recent}},
both with a logarithmic energy spectrum \cite{Mack}. 
In the {\em one}-dimensional case {\em two} interacting bosonic atoms
are excited into a state with an energy determined by
the number to be factored and the measured 
single-particle energies yield the factors. 
For the {\em two}-dimensional case a {\em single} atom suffices with the factors being
given by the two energies contained in the two
motional degrees of freedom.

In the present article we extend our method to the
factorization of an arbitrary natural number 
into an {\it a priori} unknown number of prime factors\footnote{
The distribution of the number of factors 
of an arbitrary integer was studied in \cite{Weiss}}.
Here we are using 
again a one-dimensional
trap with a logarithmic energy spectrum \cite{DiPumpo}. 
The lack of knowledge, however, leads us to a completely new
protocol of factorization.

Our article is organized as follows.
In Section \ref{Essential idea} we first introduce the logarithmic
energy spectrum and discuss the distribution of a given energy which depends on the
product of a given number of prime factors onto a collection
of single-particle states called factor states. We then employ
these results to develop
an iterative scheme which finally leads to
the prime factors. Section \ref{Model system} discusses the realization 
of our idea by bosonic atoms moving in a trap with an appropriate one-dimensional
potential. Moreover, we solve the Schr{\"o}dinger equation
that governs the dynamics of 
the bosons under the influence of a time-dependent perturbation
with\-in the rotating-wave approximation.
In Section \ref{Measurement} we first show that after a measurement
of the single-particle energies at randomly chosen
times one of the factor states is found with a probability of approximately one half.  
If, however, additional information like {\it e.g.} the {\em number}
of prime factors is available 
the factor state can be found with a probability of
about unity. 
Limitations of our method caused by the experimental
conditions are discussed in Section \ref{Limitations}.

Finally, we append a study of a different factorization scheme in Section \ref{PrimeSpect}
provided by a rather exotic potential which, nevertheless, has many
advantages over the protocol discussed before. The realization
of the potential, however, remains an open question.
We close our article in Section \ref{Summary} with a short summary while background material
and calculational details are collected in four appendices.

\section{Essential idea}\label{Essential idea}
In the present section we first introduce the logarithmic energy spectrum and
discuss its special role in finding the factors of an integer. We then turn
to the distribution of a given energy onto several subsystems.
This discussion constitutes the foundation for our iterative
factorization protocol.
\subsection{Distribution of logarithmic energy onto subsystems}\label{Distribution}
Our factorization scheme is based on a logarithmic
energy spectrum of the type
\begin{equation}\label{SinglePartSpect}
E_\ell(L) \equiv \hbar \omega_0 \ln\left(\frac{\ell}{L}+1\right),\quad \ell=0,1,2,\ldots
\end{equation}
with $E_0(L)=0$.
Here, the constant $L$ plays the role of a scaling parameter and $\hbar\omega_0$
is the unit of energy. 

Given $n$ prime numbers $p_i$ which form the vector ${\bf p}\equiv(p_1,p_2,\ldots,p_n)$
we now consider the following problem: Can an energy
\begin{equation}\label{nPartEn}
E_{\rm total}^{(n)}\equiv\hbar\omega_0\,\ln\left(\frac{\prod_{i=1}^n p_i}{L^n}\right)
\end{equation}
determined by the product of the $n$ prime factors $p_i$
be distributed onto $m$ non-vanishing energies of type (\ref{SinglePartSpect}),
and if so under which condition is this energy decomposition unique?

In order to answer this question we first note that we can cast $E_{\rm total}^{(n)}$
into the sum
\begin{equation}\label{AddTheorem}
E_{\rm total}^{(n)}=E_{p_1-L}(L)+E_{p_2-L}(L)+\ldots + E_{p_n-L}(L)\equiv E_{\bf p-L}(L)
\end{equation}
of logarithmic energies (\ref{SinglePartSpect}).
The $n$ energies on the right-hand side
belong to the spectrum (\ref{SinglePartSpect})  if the scaling parameter $L$
is positive and integer, while all energies $E_{p_i-L}$ are positive  if 
all primes $p_i$ obey $p_i > L$.  

Next we discuss the number $m$ of energies defined by (\ref{SinglePartSpect})
onto which we want to distribute the total energy $E_{\rm total}^{(n)}$.
If $m<n$ the desired distribution is possible but not in a unique way.

For example, the energy $E_{\rm total}^{(3)}(L)$ defined by (\ref{nPartEn}) 
and determined by $n=3$ prime factors
with $N=p_1\cdot p_2^2$ can
be distributed in two ways onto $m=2$ energies (\ref{SinglePartSpect}) because here
\[
E_{\rm total}^{(3)}=E_{p_1\cdot p_2-L}(L)+E_{p_2-L}(L)=E_{p_1-L}(L)+E_{p_2^2-L}(L).
\]
If, however, the number $n$ of prime factors of $N$ agrees with the number $m$ of 
logarithmic energies (\ref{SinglePartSpect}) the distribution
of $E_{\rm total}^{(n)}$ is unique because
the fundamental theorem
of arithmetics \cite{Hardy} guarantees that the prime factorization of a natural number 
is unique and so is the decomposition on the right-hand side
of (\ref{AddTheorem})\footnote{Recall that unity is defined not to be prime.
To exclude a factor unity 
from the product in (\ref{nPartEn}) 
we have to choose a scaling parameter $L\ge 2$.}.

The case $m>n$ is trivial since the product $\prod_1^n\,p_i$ cannot be
decomposed into more than $n$ factors. Obviously the same is true
for the distribution of the energy (\ref{nPartEn}). 

In  \ref{OddScaling} 
it will turn out that for our purposes we need stronger
conditions: All factors have to obey $p_i>L$. Moreover, $L$ has to be odd
which means $L\ge 3$. Together with these additional assumptions the results of the present section are the
central tools for our factorization
protocol, which we shall outline in the next section.

\subsection{Factorization protocol} \label{Protocol}

As we shall see below for every step in our iterative protocol we
need an ensemble
of $k$ identical systems each with an energy
spectrum (\ref{SinglePartSpect}) characterized by a single
quantum number $\ell\ge 0$. The integer $k$ is stepwise increased from
$2$ to $n+1$. The subsystems together form a
larger system with $k$ degrees of freedom  which constitutes the basis
for our factorization procedure.

We want to perform a complete factorization of a given integer
\begin{equation}\label{Npq}
	N \equiv \prod_{i=1}^n\,p_i,
\end{equation}
where neither the number $n$ nor the values of the prime factors $p_i$
are known.
However, to be able to  apply the results of the last section
 $N$ must not 
 contain factors $p_i\le L$.
In the case of a scaling parameter $L=3$, for example, 
this requirement is unimportant
because  factors $2$ or $3$ are easy to recognize and to remove
from the number $N$ to be factored.

To find the prime factors $p_i$ we perform a sequence
of measurements.
In the first step we 
compose a system from two subsystems
with spectrum (\ref{SinglePartSpect}) and try to prepare it in a state
with total energy 
\begin{equation}\label{twoFactors}
E_{\rm{total}}^{(2)}=\hbar\omega_0\ln\left(\frac{N}{L^2}\right).	
\end{equation}
Then we measure the energies of the two subsystems. 
This step may serve as  a primality test. Indeed, if $N$ happens to be
prime the energy (\ref{twoFactors}) can be
written as
\begin{equation}
E_{\rm total}^{(2)}=E_{N-L}+\hbar\omega_0\ln\left(\frac{1}{L}\right).
\end{equation}
The second term on the right-hand side evidently is not part of the
single-particle spectrum (\ref{SinglePartSpect}) and therefore 
$E_{\rm{total}}^{(2)}$
cannot be distributed onto the two subsystems. 
So if we fail to prepare the system with energy (\ref{twoFactors})
we have proven that $N$ is prime.

However, if $N$ is not prime
the subsystems were found with
energies $E_{m_1-L}$ and $E_{m_2-L}$, respectively. Clearly, $m_1$ and $m_2$
are factors of $N$ but we still do not know if they are prime
or composite. 

Since we want to find the {\em prime} factors
of $N$, we iterate the procedure while changing 
the number of subsystems from $k$ to $k+1$ and the total energy from
\[
E_{\rm total}^{(k)}\equiv\hbar\omega_0\ln\left(\frac{N}{L^k}\right)\quad\mbox{to}\quad
E_{\rm total}^{(k+1)}\equiv\hbar\omega_0\ln\left(\frac{N}{L^{k+1}}\right).
\]

In every step the energies 
\begin{equation}\label{EpL}
E_{m_i-L}(L)=\hbar \omega_0\,\ln\left(\frac{m_i}{L}\right),\quad i=1,\ldots,k
\end{equation}
of the $k$ subsystems
were measured and the factors $m_i$ are found by inverting (\ref{EpL}).

We continue this process until $k$ is equal to the number $n$ of prime factors.
Since $n$ is unknown
we need to find it. For this purpose we increase $k$ once more. 
However, now the new energy cannot be distributed
on the $k+1$ 
subsystems
as is shown by the expression
\begin{equation}\label{OneMore}
	\hbar\omega_0\,\ln\left(\frac{N}{L^{k+1}}\right)
	=\hbar\omega_0\left[\sum_{i=1}^n\, \ln\left(\frac{m_i}{L}\right)\right.
	+\left.\,\ln\left(\frac{1}{L}\right)\right],
\end{equation}
where the right-hand side is not a sum of energies of (\ref{SinglePartSpect}).
Hence, the $k$ factors $m_i$ found in the iteration before are the $n$ 
prime factors $p_i$ of $N$. 

We emphasize that our protocol comprises $n$ steps until the $n$ prime factors
of $N$ are found. One could think of a different protocol 
which iterates the first step
onto the two factors $m_1$ and $m_2$. Each one of them either produces 
two new factors or is identified as prime.
Generally more steps are needed than in our protocol {\it e.g.}
in the case of mutually different prime factors the number of
steps is equal to $2n-1$.

In the remainder of our  article 
we describe how the system and the
subsystems used in the protocol above are realized as well
as how the system is transferred into a state with a definite 
energy. 
Moreover, we show how to find out
if this transfer is possible or not.

\section{Model system for factorization protocol}\label{Model system}

In the preceding section we have outlined our idea of factoring an
arbitrary integer consisting of an {\it a priori}
unknown number of primes.
We now describe in detail a model system achieving this task.
For this purpose we first introduce the relevant
time-independent Schr{\"o}dinger equation of $n$ particles
each moving in a one-dimensional potential leading to a
logarithmic energy spectrum. We then couple these motions
by a time- and position-dependent interaction. The number
to be factored is encoded in the frequency of this perturbation.
The resulting infinite system of coupled equations reduces to 
a finite system which we can solve. For a brief introduction
into bosonic quantum states and the details of this
reduction we refer to \ref{BosonState} and 
B, respectively.

\subsection{One-dimensional motion of $n$ particles}\label{OneDmotion}

The subsystem with energy spectrum (\ref{SinglePartSpect}) can be
realized by a particle with mass $\mu$ which moves along the x-axis
in a one-di\-men\-sional poten\-tial $V^{(1)}=V^{(1)}(x;L)$. Its real-valued
energy wave functions $\varphi_\ell\equiv \varphi_\ell (x;L)$
obey the time-inde\-pen\-dent
Schr\"o\-din\-ger equa\-tion 
\begin{eqnarray}\label{SchrEq}
{\hat H}_0^{(1)}(x;L)\,\varphi_\ell(x;L)&\equiv& \left(-\frac{\hbar^2}{2\mu}\,\frac{d^2}{dx^2}+V^{(1)}(x;L)\right)\varphi_\ell(x;L)\\ &=&E_\ell (L)\, \varphi_\ell(x;L).\nonumber
\end{eqnarray}

For a given scaling parameter $L$ we can construct numerically the potential 
$V^{(1)}= V^{(1)}(x;L)$ such that
the solutions of (\ref{SchrEq})
just reproduce the spectrum (\ref{SinglePartSpect}) 
for the energies $E_\ell (L)$. 
Here we have indicated by the argument $L$ after the semicolon
that the potential $V^{(1)}$, as well as the wave functions $\varphi_\ell$ depend on the scaling parameter $L$.
Note that the wave functions $\varphi_\ell=\varphi_\ell(x;L)$
are even (odd) for even (odd) integer indices $\ell$.
 
Our iteration algorithm to obtain $V^{(1)}$ is
based on the Hellmann-Feynman theorem and is described
in a previous article \cite{Mack}. 
In Fig. 1  
we show $V^{(1)}$
together with the eigenfunctions $\varphi_\ell$ for $0 \le \ell \le 6$
for the case $L=3$. No degeneracy is present in
this one-dimensional problem. 

In order to realize the composite system described in Section \ref{Protocol}
we consider an ensemble of $k$ non-interacting particles each one
moving along the $x$-axis under the influence of the potential
$V^{(1)}(x;L)$. 
Using a shorthand notation ${\bf x}\equiv(x_1,x_2,\ldots,x_{k})$ for the $k$ coordinates 
$x_i$ and  ${\bf m}\equiv(m_1,m_2,\ldots,m_{k})$ 
for the $k$ quantum numbers $m_i$ of the $k$ subsystems 
the  separable solution of the 
$k$-dimensional Schr\"o\-din\-ger equation  
\begin{eqnarray}\label{SchrEqnD}
 \hat{H}^{(k)}_0({\bf x};L)\,\varphi_{\bf m}({\bf x};L)&\equiv
 &\left(\,\, \sum_{i=1}^k\,\hat{H}_0^{(1)}(x_i;L) \right)
\varphi_{\bf m}({\bf x};L)\\
&=&E_{\bf m}(L)\, \varphi_{\bf m}({\bf x},L)\nonumber
\end{eqnarray}
reads
\begin{equation}\label{phi2}
\varphi_{\bf m}({\bf x};L)\equiv \prod_{i=1}^k \varphi_{m_i}(x_i;L),
\end{equation}
with 
\begin{equation}\label{EmnD}
E_{\bf m}(L)\equiv \sum_{i=1}^k\,E_{m_i}(L).
\end{equation}  
In the next section we shall show that
the factorization protocol described 
above
can be executed with the help of the $k$-particle system (\ref{SchrEqnD})
provided we are able to prepare it
in a state with energy (\ref{EmnD}).
In the remainder of this article
we suppress 
for the sake of
simplicity in notation 
the scaling parameter $L$
in the argument of the wave functions 
and the energies.
\begin{figure}[t]
\hfill\includegraphics[height=11cm,width=11.0cm]{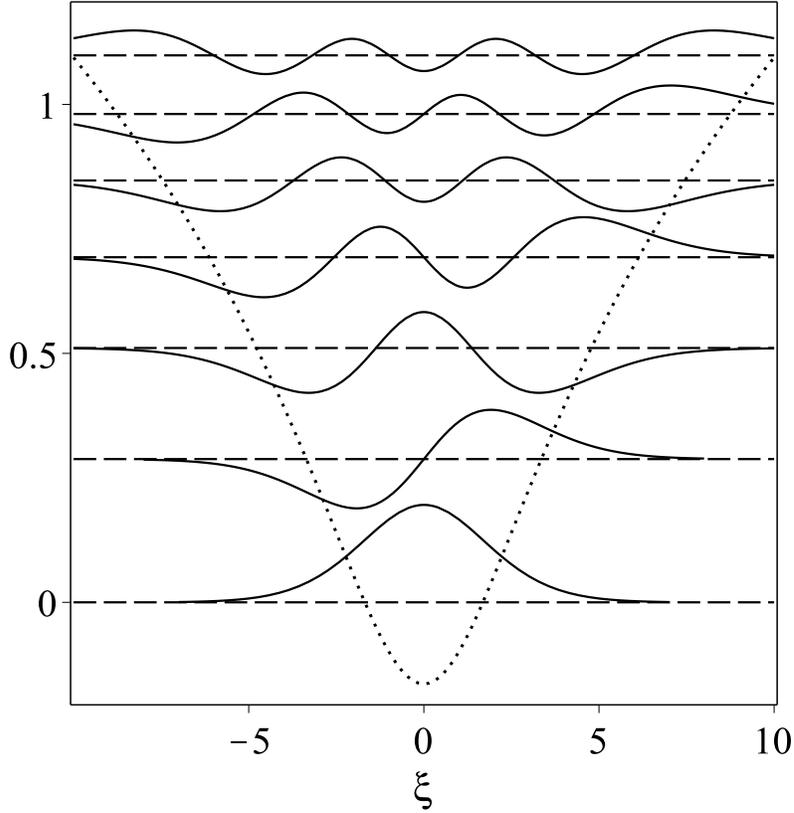}
\caption{\label{V1D}
Dimensionless one-dimensional potential 
$V^{(1)}(\xi;L=3)/\hbar\omega_0$ 
creating a logarithmic 
energy spectrum together with the corresponding wave functions for the
scaling parameter $L=3$ as a function of dimensionless coordinate 
$\xi\equiv \alpha \, x$ with $\alpha^2\equiv\mu \omega_0 /\hbar$.
This potential is determined numerically by an iteration algorithm based on a 
perturbation theory using the Hellmann-Feynman theorem, 
and is designed to obtain the logarithmic dependence of the energy 
eigenvalues $E_\ell (L)$ on the quantum number 
$\ell$ prescribed by (\ref{SinglePartSpect}).
In the neighborhood of the origin $V^{(1)}$ is approximately
harmonic whereas for large values of $\xi$ it is logarithmic. 
}
\end{figure}

\subsection{Time-dependent perturbation}\label{TimeDep}

According to the factorization protocol described in Section
\ref{Protocol} we need to
prepare the system with Hamiltonian $\hat{H}^{(k)}_0$ defined by (\ref{SchrEqnD})
in a state of total energy 
\begin{equation}\label{Etot}
E_{\rm total}^{(k)}=\hbar\omega_0\ln\left( \frac{N}{L^k}\right). 
\end{equation}

We assume that at time $t=0$ our system is in the $k$-particle
ground state $\ket{{\bf {0}}}$ 
and the perturbation 
\begin{equation}\label{Perturb}
\delta {\hat V}^{(k)}(t) \equiv\gamma^{(k)} \,\sin(\omega_{\rm{ext}}t)\,v^{(k)}({\hat {\bf x}})
\end{equation}
is switched on. The external frequency 
\begin{equation}\label{Oext}
\omega_{\rm{ext}}\equiv\omega_0 \,\ln\left( \frac{N}{L^{k}}\right)
\end{equation}
is chosen such that the energy $\hbar\omega_{\rm ext}$ agrees with $E_{\rm total}^{(k)}$
and is determined by the number $N$
we want to decompose into $k$ factors $q_j$ which may or may not be prime.

The spatial part $v^{(k)}\equiv v^{(k)}({\hat{\bf x}})$ of the interaction 
specified in \ref{SymmInt} couples all $k$ degrees of freedom. 
The strength of $\delta {\hat V}^{(k)}$ given by (\ref{Perturb}) 
is characterized by the constant $\gamma^{(k)}$.

We derive the equations of motion for the amplitudes 
$b_{\bf m}(t)$ for a transition  from the ground state $\ket{\bf 0}$
to the eigenstate $\ket{\bf m}$ by expanding
the solution $\ket{\Psi(t)}$ of the Schr\"odinger equation
\[
i\hbar\frac{d}{dt}\,\ket{\Psi(t)}=\left(\hat{H}^{(k)}_0+\,\,\delta\hat{V}^{(k)}(t)\right)\,\ket{\Psi(t)}
\]
into the eigenstates $\ket{{\bf {m}}}$ of the unperturbed Hamiltonian
$\hat{H}^{(k)}_0$,  that is
\begin{equation}\label{ExpandedPsi}
	\ket{\Psi(t)}=\sum_{\bf {m}}\, b_{\bf {m}}(t)\,\,e^{-iE_{\bf {m}}t/\hbar}\,\ket{{\bf {m}}}.
\end{equation}

Since the energies $E_{\bf m}$ given by (\ref{EmnD}) are the 
eigenvalues of ${\hat H}^{(k)}_0$ 
we arrive at the coupled system
\begin{equation}\label{CoupledS}
i \,\hbar\,{\dot b}_{\bf m}(t)=\gamma^{(k)}\sin(\omega_{\rm{ext}} t)\sum_{\bf {n}}\,
e^{i(E_{{\bf {m}}}-E_{\bf {n}})t/\hbar}\,\,W_{{\bf {m}},{\bf {n}}}\,b_{\,\bf {n}}(t)\nonumber
\end{equation}
which has to be solved for the initial conditions
\begin{equation}\label{InVal}
b_{\bf 0}(0)=1,\quad b_{\bf m}(0)=0 \,\,\, \mbox{  for } 
\sum_{i=1}^{k} \,m_i >0.
\end{equation}
The matrix elements 
$W_{{\bf m},{\bf n}}\equiv\bra{\bf m} v^{(k)}(\hat {\bf x})\ket{\bf n}$
are built from the spatial part $v^{(k)}({\hat {\bf x}})$
of the interaction (\ref{Perturb}) and the unperturbed eigenstates 
$\ket{\bf m}$. 

In the present article we consider identical bosonic atoms.
Therefore, the state vectors $\ket{\bf m}$, the probability
amplitudes $b_{\bf m}$, and the matrix elements $W_{{\bf m},{\bf n}}$
are understood as ''bosonic'' ones
in the sense defined by \ref{BosonState}.
Moreover, the summation in (\ref{ExpandedPsi}) and (\ref{CoupledS})
is performed according to (\ref{BosonSum}).
In the next section we 
show that the system under the influence of the perturbation (\ref{Perturb})
indeed can be found in a state with energy (\ref{Etot}).

\subsection{Solution of coupled equations}\label{RedSys}

In \ref{Solution} 
we use the so-called secular or rotating wave approximation (RWA) \cite{Cohen}
to reduce the infinite system (\ref{CoupledS})  
to a finite number of equations with
constant coefficients and derive the $d+1$ coupled equations 
\begin{equation}\label{bq0}
{\dot b}_{\bf 0}(t)=\frac{\gamma^{(k)}}{2\hbar}\sum_{j=1}^d\,W_{{\bf 0},{\bf q}_j-{\bf L}}\,b_{\,{\bf q}_j-{\bf L}}(t)
\end{equation}
and
\begin{equation}\label{bqj}
{\dot b}_{{\bf q}_j-{\bf L}}(t)=\frac{\gamma^{(k)}}{2\hbar}W_{{\bf q}_j-{\bf L},{\bf 0}}\, b_{\,\bf 0}(t) 
\end{equation}
with  $j=1,\ldots d$.
The degeneracy $d$ is discussed in \ref{Degeneracy}.

Equations (\ref{bq0}) and (\ref{bqj}) 
capture the essential features of the dynamics of the $k$ boson system.
Together with the initial values (\ref{InVal}) and with the help 
of the symmetry relation (\ref{Wsymm}) they lead to the solutions
\begin{eqnarray}
b_{\bf 0}(t) & = & \cos(\Omega_k\,t),\label{Solution_0}\\
b_{{\bf q}_j-{\bf L}}(t) & = & \frac{W_{{\bf 0},{\bf q}_j-{\bf L}}}{\sqrt{\sum_{i=1}^d W_{{\bf 0},{\bf q}_i-{\bf L}}^2}}\,\sin(\Omega_k\, t)\label{Solution_j}
\end{eqnarray}
with the frequency
\begin{equation}\label{Rabi}
\Omega_k \equiv \frac{\gamma^{(k)}}{2\hbar}\sqrt{\sum_{j=1}^d W_{{\bf 0},{\bf q}_j-{\bf L}}^2}\,.
\end{equation}

Equations (\ref{Solution_0}) and (\ref{Solution_j}) represent
our main result: The system of $k$ bosons performs
a Rabi oscillation, {\it i.e.} it oscillates between
the ground state $\ket{\bf 0}$ and a superposition of the $d$ factor states
$\ket{{\bf q}_i - {\bf L}}$ with {\em one single} frequency $\Omega_k$
we henceforth call Rabi frequency\footnote{The problem how to exclude a
vanishing of $\Omega_k$ is postponed to 
\ref{OddScaling}.}. 
We emphasize that the solutions 
(\ref{Solution_0}) and (\ref{Solution_j}) of the system of
differential equations (\ref{CoupledS}) are exact within the
rotating wave approximation.

The system can be found 
with probability $|b_{{\bf q}_i-{\bf L}}(t)|^2$ in
the factor state $\ket{{\bf q}_i-{\bf L}}$ 
and, at times equal to an odd multiple of $\pi/(2 \Omega_k)$ 
with certainty. Consequently, a measurement could
be performed at these times provided the Rabi frequency $\Omega_k$
would be  known.

\section{Measurement strategies}\label{Measurement}

Unfortunately, an estimate of the Rabi frequency $\Omega_k$ 
requires additional information, 
namely the knowledge of
the degeneracy $d$ which in turn depends on the number $n$ of prime
factors and on their multiplicities $\nu_i$. 
In this article we
assume that in general these parameters are unknown. 

Therefore, we now
present an alternate approach
to determine the single particle energies (\ref{SinglePartSpect}) 
of the $k$-boson system.
Moreover, we also sketch 
how a Rabi frequency can be determined if additional information
is available.

Indeed, two measurement strategies offer themselves: (i) We make a measurement
at a random time in the interval from $t=0$ where the perturbation
$\delta V^{(k)}$ is switched on, and the time $T$. The probability
to find any one of the states $\ket{{\bf q}_i-{\bf L}}$ approaches one half for $\Omega_k^{-1}\ll T$, and (ii) a measurement
at a well-defined time $t_n \equiv \pi/(2 \Omega_n)$ which requires
knowledge of the Rabi frequency $\Omega_n$ of the $n$-boson system.
Here, the
probability to find the unique factor state $\ket{{\bf p}-{\bf L}}$ is unity.  In \ref{MatrixElements} we derive an asymptotic
expression for $\Omega_n$.

\subsection{Probabilistic approach}\label{Probabilistic}

The measurement is performed at time $t$ chosen at random
from a time interval $[\,0,T]$.
According to (\ref{Solution_j}) the time-dependent probability
to find any of the $d$ factor states $\ket{{\bf q}_i-\bf{L}}$
reads 
\begin{equation}
\sum_{i=0}^d \, b_{{\bf q}_i-{\bf L}}(t_{\rm m})^2= \sin^2 (\Omega_k t_{\rm m}).
\end{equation}

\begin{figure}[t]
\hfill\includegraphics[height=7cm,width=10cm]{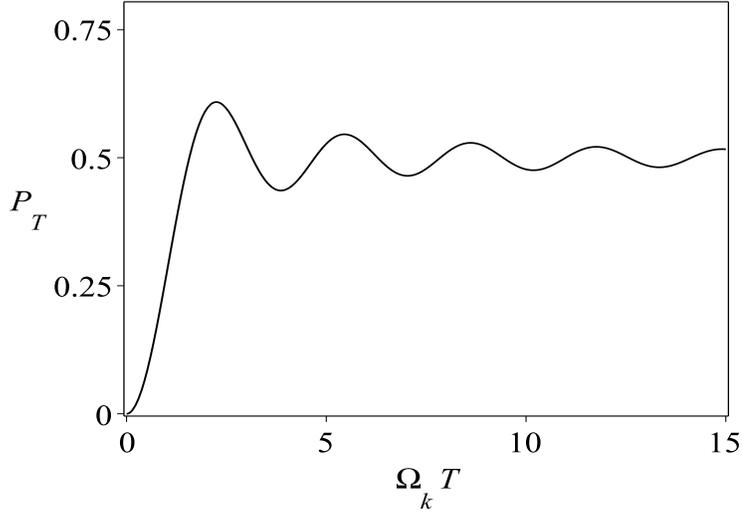}
\caption{\label{PT}
Average probability $P_T$ to find a factor state $\ket{{\bf q}_i-{\bf L}}$
at a time of measurement $t_{\rm m}$ chosen randomly from a  time interval $[0,T]$
versus dimensionless interval length $\Omega_k T$. The perturbation (\ref{Perturb})
always starts at $t=0$. For large $T$ the probability $P_T$ is about $1/2$.
}
\end{figure}
In Fig. \ref{PT} we show the average probability
\begin{equation}\label{PTeq}
P_T\equiv\langle \sin^2 (\Omega_k t) \rangle_T \equiv \frac{1}{T}\int_0^T \sin^2 (\Omega_k t)\, dt=\frac{1}{2}-\frac{\sin(2\Omega_k T)}{4 \Omega_k T}
\end{equation}
to find a factor state
as a function of the duration $T$ of the time interval. 
Evidently, the best choice for $T$ is the
decoherence time where we tacitly assume 
$T\gg \Omega_k^{-1}$.
Then $P_T\approx 1/2$ and the ground state $\ket{\bf 0}$
and any of the factor states emerge with probability $1/2$, respectively.
If a factor state $\ket{{\bf q}_j -{\bf L}}$ is found we have
succeeded, but 
if the ground state is obtained the
procedure has to be repeated until either a factor state is found, or a transition into this state is excluded
with a sufficiently large confidence interval.

We recall that the famous Shor algorithm \cite{Shor, Arendt}
can find a factor of an odd integer $N$ 
composed from $m$ different prime factors of
arbitrary multiplicity
 with a probability 
\begin{equation}
P_{\rm Shor} = 1 - \frac{1}{2^{m-1}}.
\end{equation}
We, however, find with probability of approximately $1/2$ {\em all}
$k$ factors of $N$ in a single run as long as the number of bosons $k$ does not exceed
the number $n$ of prime factors of $N$. It is amazing and amusing but
seems to be sheer coincidence
that for integers composed from two different primes 
both protocols succeed with about the same probability $1/2$.

\subsection{Additional information}\label{Estimation}

If the number $n$ of prime factors $p_i$ is known we modify
the factorization protocol of Section \ref{Protocol} which now
comprises one single step only. We use $n$ bosons in the trap
and an external frequency $\omega_{\rm ext}\equiv \omega_0 \ln (N/L^n)$ 
to excite a Rabi oscillation between the ground state $\ket{\bf 0}$ and 
the non-degenerate factor state $\ket{\bf p-L}$.
Again a random measurement results with a probability $P_T\approx 1/2$
in the single-particle energies (\ref{SinglePartSpect}) and hence
in the desired $n$ prime factors $p_i$.

If, in addition, the multiplicities $\nu_j$ of the $n$ prime factors with 
$\sum_j\nu_j=n$ are known we may estimate the Rabi frequency
\begin{equation}
\Omega_n = \frac{\gamma^{(n)}}{2\hbar}\, W_{{\bf 0},{\bf p-L}}.
\end{equation} 
The system is excited with $\omega_{\rm ext}$ as above and 
the measurement now is made not at a random time but 
at time $t_n \equiv \pi/(2\Omega_n)$. Hence, the system is found in the
factor state $\ket{\bf p-L}$ with high probability $|b_{\bf p-L}(t_n)|^2$ 
given by (\ref{Solution_j}),
and the prime factors are determined as before.

In \ref{MatrixElements} we give an example how to
calculate the Rabi frequency (\ref{Rabi}).
Here we only give the scaling properties 
\begin{equation}\label{Scaling}
\Omega_n\propto \left\{ \begin{array}{r@{\quad \quad}l} N^{-1/2} &  \mbox{for $n$ even},\\
N^{-(1/2+{(2L-1)/(2n^2)})}  &  \mbox{for $n$ odd} \end{array} \right.
\end{equation}
of the
asymptotic expression of $\Omega_n$ for large $N$
derived for  a very
simple pattern of factorization 
and refer for the prefactors and calculational details to \ref{MatrixElements}. 
We emphasize that according to (\ref{Scaling}) for an odd number $n$
of factors the Rabi frequency decreases faster with increasing $N$.

\section{Limitations}\label{Limitations}

According to Ref. \cite{Cohen} it is a shortcoming of the
RWA that transitions into off-resonant states are more probable
when the difference between these states and the
excitation energy $\hbar \omega_0 \ln(N/L^n)$ becomes comparable to or even smaller
than the energy $\hbar \Omega_n$ of the Rabi oscillation\footnote{
We tacitly assume that all transitions are allowed.}.
\setcounter{footnote}{0}
Therefore, in order to neglect transitions into off-resonant states 
we have the condition

\begin{equation}\label{OffRes}
\hbar\omega_0 \left|\,\ln\left(\frac{N\pm 1}{L^n}\right)-\ln\left(\frac{N}{L^n}\right)\right|
\approx \frac{\hbar\omega_0}{N}
\gg \hbar \Omega_n
\end{equation}
which apparently is not difficult because the
Rabi frequency $\Omega_n$ defined by (\ref{Rabi}) can be made arbitrarily small
by decreasing the strength $\gamma^{(n)}$ of the perturbation (\ref{Perturb}).

Unfortunately, a second condition arises from Section \ref{Probabilistic}
where a measurement of the energies of the $n$ bosons was performed
at times $t_m$ chosen randomly from a time interval of length $T$. To find
a factor state with probability $\approx 1/2$ requires the condition 
$\Omega_n T\gg 1$ following from 
(\ref{PTeq}).

 Clearly, the system has to be free
of decoherence during the time interval $[0,T]$ hence, the Rabi frequency $\Omega_n$
has to obey the inequality
\begin{equation}\label{gTdec}
\Omega_n\, T_{\rm dec} \, \gg 1,
\end{equation}
determined by the decoherence time $T_{\rm dec}$,
indicating that in contrast to (\ref{OffRes}) resulting from the
suppression of the off-resonant transitions $\Omega_n$
must have a lower limit as well.

To demonstrate that these conflicting conditions can be fulfilled
simultaneously
we refer to \ref{MatrixElements} where the Rabi frequency $\Omega_n$
was calculated for the simple example $N=p^n$. Having in mind that
$\Omega_n$ depends on both, the integer to be factored $N$ and the
interaction strength $\gamma^{(n)}$, we determined numerically
values of $N$ and $\gamma^{(n)}$ which cause the Rabi frequency $\Omega_n$
to obey the inequalities
\begin{figure}[t]
\hfill\includegraphics[height=10cm,width=11cm]{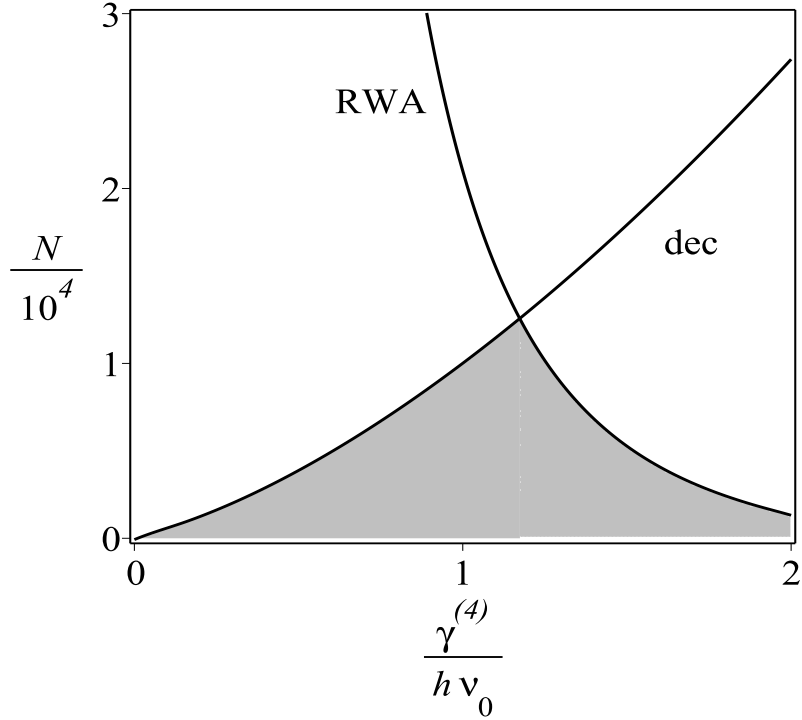}
\caption{\label{Ngamma}
Integer $N=p^4$ to be decomposed into primes as a function of
the dimensionless interaction strength $\gamma^{(4)}/(\hbar\omega_0)\equiv \gamma^{(4)}/(h \nu_0)$.
Points below the curve marked with ''RWA'' 
correspond to 
Rabi frequencies $\Omega_4$ defined by (\ref{RabiApp}) which obey the
inequality (\ref{RabiRWA}),
those below the curve marked with ''dec'' satisfy (\ref{RabiDec}).
But for points inside the shaded region the associated
Rabi frequencies enjoy {\em both}
inequalities, (\ref{RabiRWA}) and (\ref{RabiDec}), respectively.
Parameters are $L=3$, $n=4$, $T_{\rm dec}= 2 \,\,\, \rm sec$, 
$\nu_0\equiv \omega_0/(2\pi)=5\,\, {\rm kHz}$.}
\end{figure}

\begin{equation}\label{RabiRWA}
\Omega_n \le \frac{\omega_0}{N},
\end{equation}
as well as
\begin{equation}\label{RabiDec}
\Omega_n T_{\rm dec}\ge 5.
\end{equation}
The condition (\ref{RabiDec}) was motivated by  Fig. \ref{PT}
and the requirement that the average probability $P_T$ to find a
factor state has reached the asymptotic value $1/2$.

At points in the plane spanned by $\gamma^{(4)}/\hbar\omega_0$ 
and $N$ below the curve marked with ''RWA''
displayed in Fig. \ref{Ngamma}  the Rabi frequency $\Omega_4$
obeys the inequality (\ref{RabiRWA}) while at those below the curve
indicated by ''dec'' $\Omega_n$ satisfies $(\ref{RabiDec})$.
The shaded region  
consists of the points where the Rabi frequency satisfies both
conditions, (\ref{RabiRWA}) and (\ref{RabiDec}), respectively.

Our calculation was performed with the help of (\ref{AppMatrixElement}) and (\ref{RabiApp}) 
and with parameters $L=3$, $n=4$, $T_{\rm dec}= 2 \,\,\, \rm sec$, 
$\nu_0\equiv \omega_0/(2\pi)=5\,\, {\rm kHz}$.

Figure \ref{Ngamma} clearly demonstrates 
that here the maximal $N$ that can be factored with our method
is about $1.2\cdot 10^4$. 
Hence, we conjecture that the upper limit of integers with
a different prime decomposition which we factored 
with the trap parameters given above
will be of the same order  
of magnitude.

\section{Factoring with a logarithmic spectrum of primes}
\label{PrimeSpect}

In this section we propose a different factorization scheme which
uses a quite exotic potential and has advantages
compared to the protocol discussed {\it in extenso} above.
Because there are many analogies to the protocol
of Section \ref{Protocol} we keep this section short.

\subsection{Potential for logarithmic spectrum of primes}

We consider an ensemble of non-interacting bosons with a 
single-particle energy spectrum\footnote{
Tran and Bhaduri \cite{Tran} have
determined the ground state fluctuations of ensembles of bosons in a trap
with a spectrum given by (\ref{PrimeSpectrum}).}
\begin{equation}\label{PrimeSpectrum}
	E_\ell = \hbar\omega_0 \ln p_\ell,\qquad \ell=1,2,3,\ldots 
\end{equation}
that is the energy levels depend on the prime numbers 
$p_\ell$ with $p_1=2,p_2=3,p_3=5,\ldots$.
The ground state with energy  $E_0 =\hbar\omega_0\ln p_0 = 0$, for $p_0=1$ was added 
because we need it in our
factorization protocol\footnote{Recall that unity is by definition not a
prime.}. We emphasize that in contrast to the energy spectrum (\ref{SinglePartSpect})
which covers {\em all} integers $\ell$ we now restrict it solely to primes.
  
\begin{figure}[t]
\hfill\includegraphics[height=10cm,width=11cm]{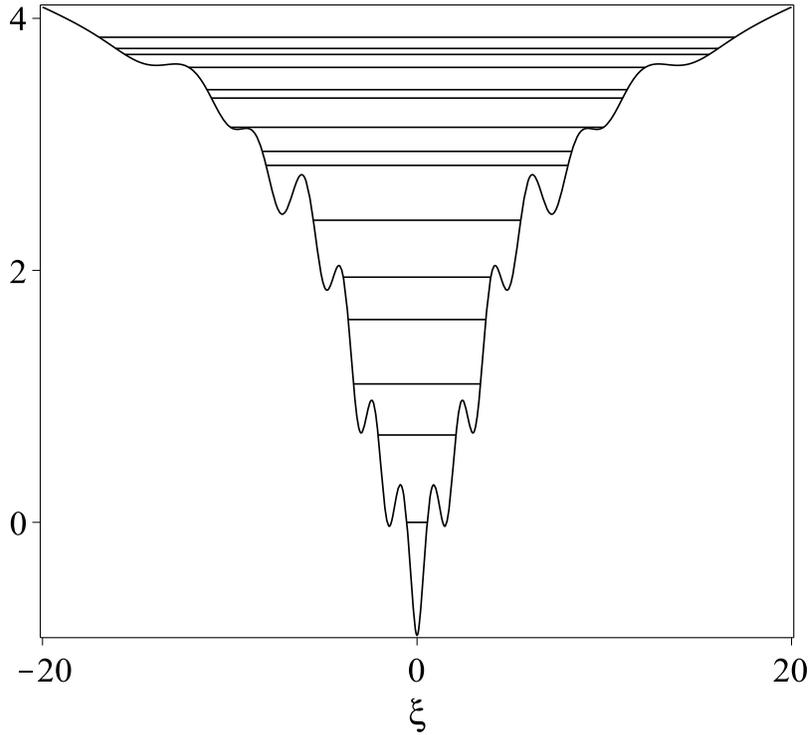}
\caption{\label{VpD}Scaled one-dimensional potential  $V_p(\xi)/\hbar\omega_0$ 
as a function of the dimensionless coordinate
$\xi\equiv \alpha\, x$ with $\alpha^2\equiv\mu \omega_0 /\hbar$
together with the corresponding energy levels.
This potential 
is designed to obtain a logarithmic dependence of the energy eigenvalues
$E_\ell$ on the {\em prime numbers} $p_\ell$ as given in Eq. (\ref{PrimeSpectrum}).
It is determined numerically by an iteration algorithm \cite{Mack} based on a 
perturbation theory using the Hell\-mann-Feyn\-man theorem. 
}
\end{figure}

We have employed methods described in \cite{Mack}
to find \cite{Wolff}
the potential $V_p=V_p(x)$ displayed in Figure \ref{VpD} together with the
$14$ lowest energy levels of the spectrum (\ref{PrimeSpectrum}).
Due to the intricate structure of $V_p$ it may be
difficult, however, to realize a trap with such a potential.

\subsection{Factorization algorithm}

Next we develop a factorization protocol analogous 
to the one of Section \ref{Protocol} but now based on the spectrum (\ref{PrimeSpectrum}). 
As before we consider an integer
\begin{equation}\label{IntegerN}
	N\equiv\prod_{i=1}^n\,p_i
\end{equation}
containing the product of $n$ unknown prime factors $p_i$. 
The number $n$ of factors is unknown as well.

From arguments already used in Section \ref{Distribution}
it follows that
an ensemble of $n$ non-interacting bosons with total energy
\begin{equation}\label{PrimeEnergy}
	E_N=\hbar\omega_0\ln N
\end{equation}
can be hosted by the spectrum (\ref{PrimeSpectrum})
in a unique way. 
But it is not possible that $m$ bosons with $m<n$ 
can be in a state with total energy (\ref{PrimeEnergy})
because in this case at least one
of them has to be in a single particle state characterized by a product of primes.
It is easy to see that, for example, an energy (\ref{PrimeEnergy}) where
the integer $N$ is
determined by $n=3$ prime factors such  as {\it e.g.} $N=p^3$ cannot
be distributed onto $m=2$ bosons with single-particle spectrum (\ref{PrimeSpectrum})
because the state $E_{N=p^3}$ is not present in the two-particle spectrum.

In contrast, an ensemble of $m$ bosons with $m>n$ can be in a state with
energy (\ref{PrimeEnergy}) because $n$ of them are in the appropriate
excited states with $E_{\ell>0}$
while the $m-n$ excess particles remain in 
the ground state of zero energy\footnote{
Note that this behavior is opposite to the one described 
in Section \ref{Protocol}  where
$m$ bosons with $m<n$
{\em can} be in a state with total energy (5)
while $m$ bosons with $m>n$ {\em cannot.}}.

Our protocol to factor the integer $N$ defined by (\ref{IntegerN}) 
is as follows. A system of $m$
bosons is prepared in the $m$-particle ground state which is even under reflection
of all position coordinates of the bosons. Like in Section \ref{TimeDep}
an adequate perturbation causes a transition into a state from which the factors of $N$
can be extracted.

Two conditions have
to be fulfilled for a successful transition: (i)
The factor state has to be even as well, and (ii) we need to have enough bosons
to cover the primes contained in $N$. We now address both conditions.

First we recall that the first excited state $\varphi_1$ with energy $E_1=\hbar\omega_0\ln 2$
is odd. Therefore, the state with energy
\begin{equation}
	E_{2N}=E_N+E_1=\hbar\omega_0\ln(2N)
\end{equation}
has opposite parity to the state with energy $E_N$.

Using $m$ bosons we choose the periodic perturbation
\begin{equation}\label{Perturb2}
\delta {\hat V}(t)\equiv \gamma^{(m)} \left\{\sin[\omega_{\rm ext}(N)\,t]+\sin[\omega_{\rm ext}(2N)\,t]\right\}\,v^{(m)}({\hat {\bf x}}),
\end{equation}
where
\begin{equation}
\omega_{\rm ext}(N) \equiv  \omega_0 \,\ln (N)\nonumber
\end{equation}
with a spatial part $v^{(m)}({\bf x})$ given by (\ref{vofx}) and {\em two} periodic terms.
Only one of them is able to cause Rabi oscillations between the ground state and a state 
which has an energy $E_N$ or $E_{2N}$, respectively. It does not matter which one of the states
is even because from any of them the factors of $N$
can be deduced as described in Section \ref{Protocol}. 

Next we turn to the second condition stating
that enough bosons have to be provided to factor $N$. While the number 2 is the smallest
prime we choose the number 
\begin{equation}\label{mfloor}
	m=\left \lfloor{\,\log_2 (N)} \right \rfloor +1
\end{equation}
of bosons
because $2 N$ can never have more than $m$ factors. 

With these conditions fulfilled
we know with certainty that a Rabi oscillation is present in the system.
Unfortunately,
we cannot estimate the Rabi frequency and we have to content ourselves
with a probability of $1/2$ to find a factor state when we measure  at an {\em arbitrary} time
the single particle energies and determine {\em all} prime factors of $N$
in a single run. However, if there
are not enough bosons in the trap no Rabi oscillation can be excited with
the perturbation (\ref{Perturb2}) and the $m$-boson 
ground state is found with a probability of 1.

\section{Summary}\label{Summary}

In the present article we have proposed two methods to find all prime factors
of an arbitrary integer $N$ based on the quantum dynamics
of a suitable number of identical bosonic atoms moving in a
one-dimensional trap. Both methods rely on exciting a collective vibratory motion
by a periodic time-dependent perturbation
whose frequency is determined by the number to be factored.
In both cases the atoms get transfered to a state where
the energies of the individual atoms represent the factors.

Our first method employs the trap
potential displayed in Fig. \ref{V1D} which 
is designed to obtain the logarithmic energy spectrum (\ref{SinglePartSpect}).
Moreover, we assume that the number $n$ of prime factors of  $N$ is unknown.
Using an iterative scheme the state characterized by the
$n$ single-particle energies is obtained after $n$ steps
where the probability  to be successful in a single run is one half. If, however,
the number $n$ of prime factors is known then this procedure is
modified such that the factor state can be obtained after one step only.

Our second method is based on the potential of Fig. \ref{VpD} where 
according to (\ref{PrimeSpectrum}) the energies of the individual atoms
depend logarithmically on the {\em prime numbers}. 
It is superior to the first one
because with enough atoms in the trap the transition into the factor state
can take place in a single step. 

Since this potential
may be difficult - if ever - to be constructed we have focused
on a thorough discussion of the first method.

\appendix

\section{Bosonic state vectors and matrix elements}\label{BosonState}

In the present appendix we briefly summarize the essential
ingredients of the quantum mechanical description of an ensemble
of bosonic atoms. Here we concentrate on the state vectors 
and the matrix elements.

\subsection{State vectors and probability amplitudes}

State vectors of $n$ identical bosonic atoms must be symmetric under
permutations of the atoms, and in the one-dimensional case
this requirement is equivalent to permutations of their quantum numbers.
The state vectors are built from product states by the well-known formula 
\begin{equation}\label{Permutations}
\ket{\bf k}_{\rm B}\equiv\ket{k_1,\ldots,k_n}_{\rm B}=
{\cal N}(\{\nu_{\bf k}\})\,\sum_{P\{{\bf k}\}}\,\ket{k_1,\ldots,k_n}.	
\end{equation}
Here the $n$ quantum numbers $k_i$ are divided into $m$ groups
containing $\nu_i$ identical numbers each with $\sum \nu_i =n$,
and the sum is over all permutations $P\{{\bf k}\}$ of the indices $k_i$
in the product states. 

The normalization factor $\cal N$ is given by \cite{Landau1977quantum}
\begin{equation}\label{Normal}
	{\cal N}(\{\nu_{\bf k}\})\equiv {\frac{\nu_1 ! \cdot \nu_2 ! \ldots\nu_m!}{n!}}
\end{equation}
since the sum in (\ref{Permutations})
contains $n!/\nu_1!,\ldots,\nu_m!\equiv {\cal N}(\{\nu_{\bf k}\})^{-2}$ terms.

Next we represent the $n$-particle state 
\begin{equation}\label{Ordinary}
\ket{\Psi(t)}\equiv {\sum_{\bf k}} b_{\bf k} (t) e^{-iE_{\bf k} t/\hbar}\ket{\bf k}.
\end{equation}
given by (\ref{ExpandedPsi})
in a basis of bosonic state
vectors $\ket{\bf k}_{\rm B}$ defined by (\ref{Permutations}).

When we sum over
bosonic states we have to avoid
counting the same state several times. Therefore, a ''bosonic'' sum 
is defined as
\begin{equation}\label{BosonSum}
	{\sum_{\bf k}}^{\rm B} \equiv \sum_{k_1=0}^\infty \, \sum_{k_2=k_1}^\infty \, \ldots\, \sum_{k_n=k_{n-1}}^\infty,
\end{equation}
that is, only summands with indices $k_1\le k_2, k_2\le k_3,\ldots k_{n-1}\le k_n$ are taken into account, and the ''ordinary'' sum $\sum_{\bf k}$ can be expressed by the identity
\begin{equation}\label{SumRelation}
\sum_{\bf k}\equiv \sum_{k_1=0}^\infty \, \sum_{k_2=0}^\infty \, \ldots\, \sum_{k_n=0}^\infty={\sum_{\bf k}}^{\rm B}\,\sum_{P\{\bf k\}}.
\end{equation}

The probability amplitudes $b_{\bf k}(t)$ are solutions of the system (\ref{CoupledS})
with coefficients $W_{\bf i,k}$ which are symmetric under the exchange of indices (\ref{Wsymm})
and, as a consequence, the amplitudes $b_{\bf k}(t)$ enjoy the same property.
With the help of (\ref{Permutations}), (\ref{Ordinary}), and (\ref{SumRelation}) it is now easy to derive the expansion
\begin{equation}\label{Bosonic}
\ket{\Psi(t)}\equiv {\sum_{\bf k}}^{\rm B} b_{\bf k}^{\rm B}(t) e^{-iE_{\bf k} t/\hbar}\ket{\bf k}_{\rm B}
\end{equation}
where the bosonic probability amplitudes are given by
\begin{equation}\label{bME}
b_{\bf k}^{\rm B}(t)={\cal N}(\{\nu_{\bf k}\})^{-1}\,b_{\bf k}(t).
\end{equation}

The square of the absolute value of the bosonic probability 
amplitude $|b_{\bf k}^{\rm B}(t)|^2$
denotes the probability $P_{\bf k}(t)$
that
after a simultaneous measurement of the energies of the bosons
performed at the time $t$
one of the bosons is found in a state with energy $E_{k_1}$,
another boson with energy $E_{k_2}$ and so on until a last one with $E_{k_n}$.

\subsection{Symmetric interaction}\label{SymmInt}

If the matrix element $A_{\bf jk}\equiv \bra{\bf j\,}{\hat A} \ket{\,\bf k}$ of an arbitrary operator $\hat A$ is symmetric under permutations of
the indices $\{j_i\}$ and $\{k_i\}$, respectively,
then the expression for the corresponding bosonic matrix element
is simply
\begin{equation} \label{bosonicME}
A_{\bf jk}^{\rm B} \equiv \left._{\rm B}\bra{\bf j\,}{\hat A} \ket{\,\bf k}_{\rm B}\right.
\end{equation}
and reads
\begin{eqnarray}
A_{\bf j,k}^B &=& {\cal N}(\{\nu_{\bf j}\})^{-1}{\cal N}(\{\nu_{\bf k}\})^{-1}\,\bra{\bf j\,}{\hat A} \ket{\,\bf k}\\
 & = & {\cal N}(\{\nu_{\bf j}\})^{-1}{\cal N}(\{\nu_{\bf k}\})^{-1}\,A_{\bf jk}.\nonumber
\end{eqnarray}

As an example of this relation
we consider the
matrix element 
\begin{equation}\label{Symm}
W_{\rm {\bf 0,k}}^{\rm B}=\left._{\rm B}\bra{\bf 0}v^{(n)}({\hat {\bf x}})\ket{\bf k}_{\rm B}\right.
\end{equation}
for a short-range collective $n$-particle interaction 
\begin{equation}\label{vofx}
	v^{(n)}({\bf x})= \alpha^{1-n} \,\prod_{j=2}^n \, \delta(x_{j-1}-x_j),
\end{equation}
which couples all
degrees of freedom of the $n$ bosons and is symmetric, that is,
$v^{(n)}({\bf x})=v^{(n)}({\bf -x})$.
Here we have introduced powers of
\begin{equation}
\alpha \equiv \sqrt{\mu\omega_0 /\hbar}
\end{equation}
as to make the matrix element $W_{\rm {\bf 0,k}}$
dimensionless.

Using the real-valued product states $\varphi_{\bf j}({\bf x})$ defined by (\ref{phi2})
and the interaction (\ref{vofx}) it is easy to derive the expression
\begin{equation}\label{W00uv}
W_{{\bf 0},{\bf q}_j-{\bf L}}^{\rm B}={\cal N}(\{\nu_{{\bf q}_j}\})^{-1}\,\int dx\,\varphi_{0}(x)^n
	\prod_{i=1}^n\,\varphi_{({\bf q}_j)_i-L}(x).
\end{equation}

Evidently, the matrix element $W^{\bf B}_{{\bf 0},{\bf q}_j-{\bf L}}$ is symmetric under any permutation of the indices 
$\{0_\ell,({\bf q}_j)_i-L\}$, and hence
\begin{equation}\label{Wsymm}
W_{{\bf q}_j-{\bf L},{\bf 0}}^{\rm B}=W_{{\bf 0},{\bf q}_j-{\bf L}}^{\rm B}.
\end{equation} 

The sub- or superscript B used here is omitted in 
the main body of our article because we only deal with bosons.

\subsection{Odd scaling parameter}\label{OddScaling}

We note that the matrix element vanishes if the product in (\ref{W00uv}) contains an odd number $n$ of 
odd wave functions $\varphi_{({\bf q}_j)_i-L}(x)$.
Since this feature emerges for every ${\bf q}_j$ with $j=1\ldots d$ the Rabi frequency $\Omega_n$ 
defined by (\ref{Rabi})
also vanishes:
a transition from the even ground state into an odd factor state cannot
take place through the interaction (\ref{vofx}). 

We recall that in 
Section \ref{Distribution} we have postulated
that the scaling factor $L\ge 3$ has to be odd. Moreover, all
prime factors obey $p_j>L$. Then any factor $q_i$, prime or composite,
is odd and consequently the difference $({\bf q}_j)_i-L$ is {\em even} and 
the integrand in (\ref{W00uv}) is even as well: 
$W_{{\bf 0},{\bf q}_j-{\bf L}}$ never vanishes 
whatever factors $({\bf q}_j)_i>L$
appear in $N$ if we have chosen an odd scaling parameter $L$
in the single-particle spectrum (\ref{SinglePartSpect}).

\section{Reduction of the coupled system of equations}\label{Solution}

In this appendix we reduce the infinite coupled system of equations (\ref{CoupledS})
to a finite one with constant  coefficients. 
This simplification is possible due to the special choice
of the logarithmic energy spectrum and the application of the rotating wave approximation.

We consider $k$ bosons with the single-particle spectrum (\ref{SinglePartSpect})
and assume first that $k$ is
less or equal to the number $n$ of prime factors $p_j$ of $N$
defined by (\ref{Npq}). Here we represent $N$ as a product
\begin{equation}\label{Nq}
N=\prod_{j=1}^k \, q_j,
\end{equation}
of $k$ factors $q_j$ which may or may not be prime. 

We substitute the resonance condition (\ref{Oext})
into the term with all $k$ integers $m_i=0$ of (\ref{CoupledS}), and after little algebra we
arrive at the relation
\begin{eqnarray}
\lefteqn{i\hbar{\dot b}_{\bf 0}(t)=\frac{\gamma^{(k)}}{2i}\sum_{\bf n}\,W_{{\bf 0},{\bf n}}
\left\{\exp\left[i\ln\left(\frac{N}{\prod_{j=1}^k (n_j+L)}\right)\omega_0 t\right]\right.}\hspace{0.7cm}\nonumber\\
      & &\left. -\exp\left[-i\ln\left(\frac{N}{L^k}\frac{\prod_{j=1}^k\,(n_j+L)}{L^k}\right)\omega_0 t\right] \right\}b_{\,\bf n}(t).\label{b001}
\end{eqnarray}
The first phase factor in the curly brackets of (\ref{b001}) is unity 
for $n_j=q_j-L$,  and for any permutation  of the factors $\{q_j\}$. 
However, according to (\ref{BosonSum}) 
only the permutation with $q_1 \le q_2 \le q_3 \ldots \le q_k$
needs to be taken into account. In the second phase factor in the
curly brackets no terms survive because for $N>L^k$ the argument of the
lo\-ga\-rithm can never be unity.

As long as $k<n$ the representation 
(\ref{Nq}) of $N$
may have more than one solution $\{q_1,q_2,\ldots q_k\}$. If, for example,
$N$ has $n=3$ prime factors which have to be dis\-tri\-bu\-ted
onto $k=2$ bosons, as $N=p_1 \cdot p_2^{\,2}$, two so\-lu\-tions 
${\bf q}_1=(p_1,p_2^{\,2})$ and ${\bf q}_2=(p_1 \cdot p_2,p_2)$ exist. 

However, if $k=n$ the representation (\ref{Nq}) has 
a unique so\-lu\-tion and all the factors 
$q_j=p_j$ are prime as follows from the fundamental theo\-rem dis\-cus\-sed
in Section \ref{Distribution}. 
Thus, we are left with the equation 
\begin{equation}\label{b001const}
{\dot b}_{\bf 0}(t)=-\frac{\gamma^{(k)}}{2\hbar}\sum_{j=1}^d\,W_{{\bf 0},{\bf q}_j-{\bf L}}\,b_{\,{\bf q}_j-{\bf L}}(t),
\end{equation}
where $d$ denotes the number of solutions ${\bf q}_j$ of (\ref{Nq}),
that is, the degeneracy
discussed in \ref{Degeneracy}.

Finally, for $k>n$ all phase factors in the curly brackets of the right-hand 
side of (\ref{b001}) oscillate and average out
according to the rotating wave approximation.
Consequently, the equation
\begin{equation}\label{b002}
{\dot b}_{\bf 0}(t)=0
\end{equation}
together with the initial conditions (\ref{InVal})
immediately yields $b_{\bf 0}(t)\equiv b_{\bf 0}(0)$, {\it i.e.} the $k$
bosons remain in their ground state $\ket{\bf 0}$.

In the remainder of this article we concentrate on the case $k\le n$.

Next we address  the system (\ref{CoupledS}) with ${\bf m}={\bf q}_j-{\bf L}$ and derive for every solution ${\bf q}_j$ of (\ref{Nq})
the equation  
\begin{eqnarray}
\lefteqn{i\hbar{\dot b}_{{\bf q}_j-{\bf L}}(t)=\frac{\gamma^{(k)}}{2i}
\sum_{\bf n}\, W_{{\bf q}_j-{\bf L},{\bf n}}\times}\hspace{0.8cm}\nonumber\\
& & {}\times\left\{\exp\left[i\ln\left(\frac{N^2}{L^k\prod_{j=1}^k (n_j+L)}\right)\omega_0 t\right]\right.\nonumber\\
 & &  \left.{}-\exp\left[-i\ln\left(\frac{\prod_{j=1}^k (n_j+L)}{L^k}\right)\omega_0 t\right] \right\}b_{\,\bf n}(t)\label{buv00}.
\end{eqnarray}
The first phase factor in curly brackets always oscillates since
the integer $N^2$ in the 
numerator of the argument of the logarithm is built from $2n$ prime
factors where each one obeys\footnote{The first phase factor would survive
for some integers $k_j$ if at least half of the factors $p_j$ were
equal to the scaling parameter $L$.}
the condition $p_j>L$
as required in Section \ref{Distribution}.

The second phase factor is unity for all indices $n_j=0$ 
leading to $d$ differential equations
\begin{equation}\label{bpq2}
{\dot b}_{{\bf q}_j-{\bf L}}(t)=\frac{\gamma^{(k)}}{2\hbar}W_{{\bf q}_j-{\bf L},{\bf 0}}\, b_{\,\bf 0}(t), \quad j=1,\ldots d.
\end{equation}
Equations (\ref{b001const}) and (\ref{bpq2}) describe the main features of
quantum dynamics induced by the time- and position-dependent perturbation $\delta V^{(k)}$.
In Section \ref{Model system} we solve these equations.

\section{Degeneracy}\label{Degeneracy}

The integer $d$ introduced in Section \ref{RedSys} denotes the number of
ways the equation
\begin{equation}\label{Neq}
N=\prod_{i=1}^n\, p_i = \prod_{j=1}^k\,q_j
\end{equation}
can be solved for $k$ positive integers $q_j$ regardless of their order.
In (\ref{Neq}) the number $N$ is built from $n$ prime factor $p_i$. Of course,
for $k=n$ the factors $q_j$ are identical to the prime factors $p_i$
due to the fundamental theorem of arithmetics, hence $d=1$
as discussed in section \ref{Distribution}.
Unfortunately, for $k\ne n$ a closed-form expression for $d$ exists only in two situations.

In the first case all $p_j$ are mutually different and
we have \cite{NIST:DLMF}
\begin{equation}\label{Stirling}
d=S(n,k)
\end{equation}
where $S(n,k)$ denotes the Stirling number of the second kind, that is
the number of partitions of a set of $n$ distinct elements 
into exactly $k$ nonempty subsets.

In the second case, if all $p_i$ are equal {\it i.e.} $N=p^n$, we have \cite{NIST:DLMF} 
\begin{equation}\label{Partition}
d=p_k (n)- p_{k-1}(n)
\end{equation}
where the restricted partition $p_k(n)$ denotes the number of
partitions of $n$ into {\em at most} $k$ parts. 

In all other cases $d$ has to be constructed ''by hand''.

\section{Matrix elements}\label{MatrixElements}

In this appendix we provide an elementary example for estimating a Rabi frequency
by considering the case where all $n$ non-interacting bosons are
in the same single-particle state $\ket{p-L}$ with $p^n=N$ and
${\cal N} (\{\nu_{\bf p}\})=1$. Now the matrix element (\ref{W00uv})
is given by 
\begin{equation}\label{W00uvpn}
W_{\bf 0,p-L}=\int dx \varphi_0(x)^n \varphi_{p-L}(x)^n.
\end{equation}
For the wave functions appearing
in the integrand 
we use the following approximations:

(i) We replace the ground state wave function $\varphi_0$ by that of
a harmonic oscillator 
\begin{equation}
	\varphi_0(x)\approx \alpha_{\rm eff}^{1/2}\left(\frac{1}{\pi}\right)^{1/4}  \, e^{-\alpha_{\rm eff}^2 x^2 / 2}
\end{equation}
with an 
inverse characteristic length 
$\alpha_{\rm eff}\equiv (\mu\omega_{\rm eff}/\hbar)^{1/2}$.
As detailed in \cite{Mack} close to the origin the potential $V^{(1)}(x)$ is 
approximately that of an harmonic oscillator with an effective
frequency 
\begin{equation}
	\omega_{\rm eff}\equiv \frac{\omega_0}{L-1/2}.
\end{equation}
Here we have changed the notation of \cite{Mack} to conform to Eq. (\ref{SinglePartSpect}) of the present article. 

(ii) For the even wave function $\varphi_\ell$
we use the simplified
WKB approximation \cite{Gleisberg,Schleich} 
\[
\varphi_\ell(x)\approx \left(\frac{2\alpha}{\pi L}\right)^{1/2}
\frac{\cos[\,\beta_\ell(L) \,\alpha x]}{(\ell/L+1)^{1/2}\,\beta_\ell(L)^{1/2}}
\] 
valid near the center 
of the trap
where 
\begin{equation}\label{betaL}
\beta_\ell(L)\equiv [\,2\ln(\ell/L+1)-2V^{(1)}(0)\,]^{1/2}.
\end{equation}

Under these assumptions, and aside from a prefactor, the right-hand side of 
the matrix element $W_{\bf 0, p-L}$ given by (\ref{W00uvpn}) is transformed into the integral
\begin{equation}\label{IntegralLeft}
	W_n\equiv \frac{1}{\sqrt \pi}\int_{-\infty}^\infty d\eta \, e^{-\eta^2}\cos^n ({\bar \beta}\eta),
\end{equation}
where we have introduced the variable
\begin{equation}
\eta\equiv \left( \frac{n}{2L-1} \right)^{1/2}\,\alpha x
\end{equation}
and the constant 
\begin{equation}\label{BetaBar}
	{\bar \beta}\equiv  \left[2\ln\left(\frac{N^{1/n}}{L}\right)\,\frac{2L-1}{n}\right]^{1/2}.
\end{equation}

For  even $n$ we use the expansion
\begin{equation}
\cos^n(x)=\sum_{j=0}^{n/2}\, d_j(n) \cos(2jx).
\end{equation}
to convert (\ref{IntegralLeft}) into a sum of standard integrals
\begin{equation}\label{StandInt}
	\frac{1}{\sqrt \pi}\int_{-\infty}^\infty e^{-\eta^2}\cos(2j{\bar \beta}\eta)=e^{-{\bar \beta}^2 j^2}
\end{equation}
while the coefficients $d_j(n)$ can be found  {\it e.g.} in \cite{Prudnikov}. 

Equation (\ref{StandInt})
shows that asymptotically for large $N$ {\it  i.e.} for large $\bar \beta$ only
the term with $j=0$ has to be taken into account.

With the help of the relation
\begin{equation}
	d_0(n)=\frac{1}{2^n}\,{ n \choose n/2 }
\end{equation}
it is a straightforward but lengthy algebra to derive the result
\begin{equation}\label{AppMatrixElement}
	W_{{\bf 0},{\bf p-L}}^{\rm even}\approx  \frac{w(L,n)\,d_0(n)}{[\,\ln(N/L^n)\,]^{n/4}}\,N^{-1/2}, 
\end{equation}
with
\begin{equation}\nonumber
  w(L,n)\equiv \left(\frac{2}{\pi}\right)^{n/2}
	\left(\frac{n}{\pi(2L-1)}\right)^{(n-2)/4} 
\end{equation}

For an  odd number of factors $n$ a similar procedure leads to 
\begin{equation}\label{AppMatrixElodd}
	W_{{\bf 0},{\bf p-L}}^{\rm odd}\approx  \frac{2\,w(L,n)\, d_0(n+1)}{[\,\ln(N/L^n)\,]^{n/4}\,(N/L^n)^{(2L-1)/(2n^2)}}\,N^{-1/2}.
\end{equation}
Because here the factor state $\ket{\bf p-L}$ is non-degenerate the Rabi
frequency 
\begin{equation}\label{RabiApp}
\Omega_n=\frac{\gamma^{(n)}}{2\hbar}\, W_{{\bf 0},{\bf p-L}}
\end{equation}
is readily determined
for both cases, even and odd, respectively.

\ack

We thank  M. G. Boshier, M. Freyberger and K. Vogel for stimulating discussions on this topic.
WPS is grateful to the Hagler Institute for Advanced Study at Texas A\&M University for a
Faculty Fellowship and to Texas A\&M University AgriLife Research for its support.
The research of the IQST is financially supported by the Ministry of 
Science, Research and Arts Baden-W{\"u}rttemberg.

\section*{References}
\providecommand{\newblock}{}


\begin{thebibliography}{10}
	\expandafter\ifx\csname url\endcsname\relax
	\def\url#1{{\tt #1}}\fi
	\expandafter\ifx\csname urlprefix\endcsname\relax\def\urlprefix{URL }\fi
	\providecommand{\eprint}[2][]{\url{#2}}
	
	\bibitem{RSA}
	Rivest R~L, Shamir A and Adleman L 1978 {\em Commun. ACM\/} {\bf 21} 120--126
	
	\bibitem{Shor}
	Shor P~W 1994 Algorithms for quantum computation: Discrete logarithms and
	factoring {\em Foundations of Computer Science, 1994 Proceedings., 35th
		Annual Symposium on\/} (IEEE) pp 124--134
	
	\bibitem{Gleisberg}
	Gleisberg F, Mack R, Vogel K and Schleich W~P 2013 {\em New Journal of
		Physics\/} {\bf 15} 023037
	
	\bibitem{Gleisberg20152556}
	Gleisberg F, Volpp M and Schleich W~P 2015 {\em Physics Letters A\/} {\bf 379}
	2556 -- 2560
	
	\bibitem{Boshier}
	{Amico} L and {Boshier} M~G 2015 {\em ArXiv e-prints\/} (\textit{Preprint}
	\eprint{1511.07215})
	
	\bibitem{Henderson}
	Henderson K, Ryu C, MacCormick C and Boshier M~G 2009 {\em New Journal of
		Physics\/} {\bf 11} 043030
	
	\bibitem{garraway2016recent}
	Garraway B~M and Perrin H 2016 {\em Journal of Physics B: Atomic, Molecular and
		Optical Physics\/} {\bf 49} 172001
	
	\bibitem{Mack}
	Mack R, Dahl J~P, Moya-Cessa H, Strunz W, Walser R and Schleich W~P 2010 {\em
		Physical Review A\/} {\bf 82} 032119
	
	\bibitem{Weiss}
	Weiss C, Page S and Holthaus M 2004 {\em Physica A\/} {\bf 341} 586--606
	
	\bibitem{DiPumpo}
	Di~Pumpo F 2015 Bachelor's thesis, {P}rimzahlzerlegung mit {H}ilfe eines
	{P}otentials mit loga\-rithmischem {E}nergiespektrum, {U}niversit\"at {U}lm
	
	\bibitem{Hardy}
	Hardy G~H and Wright E~M 1979 {\em An Introduction to the Theory of Numbers 5th
		edition\/} (Oxford University Press)
	
	\bibitem{Cohen}
	Cohen-Tannoudji C, Diu B and Lalo{\"e} F 1977 {\em Quantum {M}echanics\/}
	(Wiley {I}nterscience, New York)
	
	\bibitem{Arendt}
	Arendt W and Schleich W~P 2009 {\em Mathematical Analysis of Evolution,
		Information, and Complexity\/} (John Wiley \& Sons, New York)
	
	\bibitem{Tran}
	Tran M~N and Bhaduri R~K 2003 {\em Physical Review E\/} {\bf 68} 026105
	
	\bibitem{Wolff}
	Wolff G 2013 Bachelor's thesis, {P}otential mit einem aus den {L}ogarithmen der
	{P}rimzahlen gebildeten {E}nergiespektrum, {U}niversit\"at {U}lm
	
	\bibitem{Landau1977quantum}
	Landau L~D and Lifshitz E~M 1977 {\em Quantum {M}echanics: {N}on-{R}elativistic
		{T}heory, \S 64\/} (Oxford New York: Pergamon Press)
	
	\bibitem{NIST:DLMF}
	{\it NIST Digital Library of Mathematical Functions,} http://dlmf.nist.gov,
	Release 1.0.13 of 2016-09-16 f.~W.~J. Olver, A.~B. {Olde Daalhuis}, D.~W.
	Lozier, B.~I. Schneider, R.~F. Boisvert, C.~W. Clark, B.~R. Miller and B.~V.
	Saunders, eds.
	
	\bibitem{Schleich}
	Schleich W~P 2001 {\em Quantum {O}ptics in {P}hase {S}pace\/} (VCH-Wiley,
	Weinheim)
	
	\bibitem{Prudnikov}
	Prudnikov A~P, Brychkov Y~A and Marichev O~I 1998 {\em Integrals and Series:
		Volume 1: Elementary Functions\/} (Gordon and Breach Science Publishers, New
	York)
	
\end{thebibliography}
\end{document}